\documentclass[sigconf,natbib]{acmart}
\usepackage[boxed]{algorithm2e}
\usepackage{hyperref}
\usepackage{soul,balance}
\usepackage{graphicx}
\usepackage{float}
\AtBeginDocument{%
  \providecommand\BibTeX{{%
    \normalfont B\kern-0.5em{\scshape i\kern-0.25em b}\kern-0.8em\TeX}}}

\setcopyright{rightsretained}

\begin{document}

\title[Building K-Anonymous User Cohorts with Consecutive Consistent Weighted Sampling (CCWS)]{Building K-Anonymous User Cohorts with\\ Consecutive Consistent Weighted Sampling (CCWS)
}

\author{Xinyi Zheng, Weijie Zhao, Xiaoyun Li, Ping Li}
 \affiliation{%
   \institution{LinkedIn Ads}
\{cazheng, weijzhao, xiaoyli, pinli\}@linkedin.com
\vspace{0.3in}
   \country{}
 }

\renewcommand{\shortauthors}{Xinyi Zhao, Weijie Zhao, Xiaoyun Li and Ping Li}

\begin{abstract}
To retrieve personalized campaigns and creatives while protecting user privacy, digital advertising is shifting from member-based identity to cohort-based identity. Under such identity regime, an accurate and efficient cohort building algorithm is desired to group users with similar characteristics. In this paper, we propose a scalable $K$-anonymous cohort building algorithm called {\em consecutive consistent weighted sampling} (CCWS). The proposed method combines the spirit of the ($p$-powered) consistent weighted sampling and hierarchical clustering, so that the $K$-anonymity is ensured by enforcing a lower bound on the size of  cohorts. Evaluations on a LinkedIn dataset consisting of $>70$M  users and ads campaigns demonstrate that CCWS achieves substantial improvements over several hashing-based methods including sign random projections (SignRP), minwise hashing (MinHash), as well as the vanilla CWS.

%could achieve 17\% to 230\% utility improvement without compromising user privacy.
\end{abstract}






\maketitle

\section{Introduction} \label{sec:intro}

The retrieval of effective advertising campaigns and creatives, as a crucial step for digital advertising, has been profoundly influenced by the data privacy policies in recent years.
Major players are taking steps to improve user privacy in the digital advertising world. For instance, Apple recently released its App Tracking Transparency (ATT) feature, which requires app developers to obtain user consent before tracking user data across apps and websites~\cite{att2021}. Google has announced its privacy sandbox initiative on chrome and Android, which limits conversion tracking~\cite{sandbox2023}. Essentially, advertising platforms cannot access device’s identifier and freely link/aggregate user data across applications and websites. In this paper, we report our solution based on a scalable \textit{cohort} construction.
Our proposed cohort-building algorithm is  built upon the technique called {\em consistent weighted sampling} (CWS)~\citep{manasse2010consistent,ioffe2010improved,li2017tunable,li2017theory}, which is a generalization of {\em minwise hashing} (MinHash) for binary data~\cite{broder1997syntactic,broder1997resemblance,broder1998min,li2005using, shrivastava2015asymmetric,li2022c}. For the purpose of comparison, a natural baseline would be the {\em sign random projections} (SignRP), or more generally quantized random projections~\cite{goemans1995improved,charikar2002similarity,datar2004locality,li2014coding, li2019random, li2021quantization, li2022signrff}.

\section{Problem Statement and Baselines}

\vspace{0.1in}
\noindent\textbf{Cohort building.} To achieve the balance between user privacy and personalization, digital advertising is shifting from member-based identity to cohort-based identity. For example, Google's privacy sandbox includes a range of privacy-enhancing technologies such as Federated Learning of Cohorts (FLoC) and FLEDGE~\cite{FLoC2020, Fledge2021}. In short, a cohort is a group of users sharing some similar characteristics. Each cohort has a cohort identity. When an ads campaign (which targets at some specific user identities) is created, cohorts whose identities match the campaign are considered as the campaign's audience. Later when there is an impression opportunity, the cohort identity of the impression is resolved, and relevant campaigns are retrieved as candidate campaigns in ranking.
There are two main advantages of this approach: (a) Advertising platforms can track user conversion based on cohort identity to prevent the first and third party data join on the member level; (b) Advertising platforms can utilize FloC and FLEDGE to continue interest-based and personalized advertising. More formally, the adoption of user cohorts is naturally in accordance with the following  requirement.

\begin{definition}[\cite{sweeney2002k,machanavajjhala2007l}]
  A cohort is \textbf{\textit{$K$-anonymous}} if it is shared by at least $K$ users, and each user cannot be distinguished from the remaining users within the cohort.
\end{definition}

In our problem, building cohorts with $K$-anonymity is fundamentally a clustering problem with a strict lower bound on the cluster size~\cite{byun2007efficient}. Therefore, for industry-scale applications with a large number of data points (e.g., $n\approx 100M$ users), two main challenges are: (a) the efficiency/scalability of the clustering algorithm; (b) the requirement on the minimal group size. In this regard, conventional clustering methods, including density-based, centroid-based, and connectivity-based algorithms, are not ideal  for solving the cohort building problem. Density-based model such as DBSCAN~\cite{ester1996density} has $O(n^2)$ time complexity
and $O(n^2)$ memory complexity (for matrix-based fast implementation) which is  expensive for large $n$ in practice. Centroid-based model such as Lloyd's $k$-means~\cite{lloyd1982least} has time complexity $O(nkI)$, where $k$ is the number of clusters and $I$ is the number of iterations. Since in applications the lower bound on the group size is often small  (such as 20), the $k$ typically has to be very large (i.e., $k=\Theta (n)$) and thus the complexity is also roughly $O(n^2)$. Moreover, both DBSCAN and $k$-means cannot guarantee a minimal cluster size. Connectivity-based model such as hierarchical agglomerative clustering~\cite{ward1963hierarchical} can enforce a cluster size lower bound, but incurs $O(n^2)\sim O(n^3)$ time and memory costs. In summary, we are in need of a good cohort building technique which is accurate, scalable, and   $K$-anonymous, at the same time.

\vspace{0.1in}
\noindent\textbf{Sign random projections (SignRP, a.k.a. SimHash).} The method of quantized random projections~\cite{goemans1995improved,charikar2002similarity,datar2004locality,li2014coding, li2019random, li2021quantization, li2022signrff} provides an effective hashing strategy for indexing, storage,  feature compression, etc. Given two data vectors $x$ and $y$, the basic idea of random projections is to compute the inner product between each data vector and a random vector whose entries are  sampled from Gaussian. The inner products are quantized to save space and provide indexing. In the extreme, we only use the signs of the random projections, which still preserve similarities between data vectors. For each data vector, we must repeat SignRP to generate multiple bits. The number of necessary repetitions depends on  applications.

\vspace{0.1in}
\noindent\textbf{Minwise hashing (MinHash).}  (b-bit) Minwise hashing is a standard hashing algorithm for the binary Jaccard similarity defined as $J(x,y)=\frac{|x\cap y|}{|x\cup y|}$ for two sets $x$ and $y$. Let $\pi:[D]\rightarrow[D]$, be a permutation mapping, where $D$ is the size of the universe. The MinHash sample is computed by $h(x)=\min(\pi(x))$. Applying the same $\pi$ to $x$ and $y$,
the MinHash collision probability  is known as
\begin{align}
    Pr(h(x)=h(y))=J, \label{eqn:collision-minhash}
\end{align}
which suggests an unbiased Jaccard estimator by generating multiple independent permutations and averaging over the corresponding collision indicators.
The Jaccard similarity and minwise hashing have been heavily used by practitioners in numerous applications~\citep{broder1997syntactic,broder1998min,fetterly2003large,nitin2008opinion,buehrer2008scalable,urvoy2008tracking,dourisboure2009extraction,forman2009efficient,pandey2009nearest,cherkasova2009applying,chierichetti2009compressing,gollapudi2009axiomatic,najork2009less,bendersky2009finding,li2011learning,shrivastava2012fast,schubert2014signitrend,shrivastava2014improved,fu2015design,pewny2015cross,manzoor2016fast,raff2017alternative,tymoshenko2018cross,zhu2019josie,lei2020locality,thomas2020preserving}. For example,  the integer hash values can be used as indexing for approximate nearest neighbor search~\cite{indyk1998approximate,shrivastava2012fast}. Basically, the first step is to generate hash values for all the data points and build a hash table, where each data point is then assigned to a bucket in the  table. By Eq.~\eqref{eqn:collision-minhash}, data points with high similarities are more likely to be landing in the same bucket. When searching for nearest neighbors, one may simply examine the bucket that the query belongs to, instead of scanning the entire database, which substantially improves the efficiency. \cite{shrivastava2014defense} showed the advantage of MinHash over SignRP for approximate nearest neighbor search on sparse data.

\vspace{0.1in}
\noindent\textbf{$p$-Powered consistent weighted sampling (CWS).} The binary Jaccard similarity can be naturally extended to general non-negative data. Given two non-negative vectors $x,y\in\mathbb R^d_+$, the weighted Jaccard similarity is  $J(x,y)=\frac{\sum_i\min(x_i,y_i)}{\sum_i\max(x_i,y_i)}$. The non-binary weights are  more informative and may lead to better performance in subsequent tasks. Analogous to the binary Jaccard, the weighted Jaccard  has been studied and used in many areas including theory, databases, machine learning, and information retrieval~\cite{kleinberg1999approximation,charikar2002similarity,gollapudi2009axiomatic,fetterly2003large,schubert2014signitrend,fu2015design,pewny2015cross,manzoor2016fast,raff2017alternative,tymoshenko2018cross,zhu2019josie,lei2020locality,thomas2020preserving,ioffe2010improved,manasse2010consistent,bollegala11a2011web,delgado2014data,wang2014learning,li2017theory, ertl2018bagminhash,bag2019efficient,pouget2019variance,yang2019nodesketch,fuchs2020intent,thomas2020preserving,li2021consistent}. It has been shown that $J(x,y)$ defines a positive-definite (non-linear) kernel, and can be further generalized to the following $p$-powered generalized min-max (pGMM) kernel~\cite{li2017tunable,li2017theory,li2022gcwsnet}:
\begin{align} \label{def:pGMM}
    pGMM(x,y;p) = \frac{\sum_i\min(x_i,y_i)^p}{\sum_i\max(x_i,y_i)^p},
\end{align}
which introduces a tuning parameter $p$ to the weighted Jaccard similarity. Note that when the data vectors have negative entries, we need to first double the data dimensions to obtain  new vectors of non-negative entries only~\cite{li2017tunable,li2017theory} (hence the name ``generalized min-max (GMM)''). Recent works have found that the pGMM kernel can outperform the popular Gaussian kernel on many tasks~\cite{li2017tunable,li2017theory, li2022gcwsnet,li2019re}. The consistent weighted sampling (CWS), summarized in Algorithm~\ref{alg:cws},  is a popular hashing method for the pGMM kernel.

\begin{algorithm}[h!]
\LinesNumbered
% \SetKwData{Left}{left}\SetKwData{This}{this}\SetKwData{Up}{up}
% \SetKwFunction{Union}{Union}\SetKwFunction{FindCompress}{FindCompress}
%\SetKwInOut{Input}{Input}
\SetKwInOut{Output}{Output}
\SetKwInput{Input}{Input}
\Input{User feature vector $u$ ($u\in\mathbb{R}^d$); pGMM kernel tunable parameter $p$; random seed $s$}
% \If{$r$ is available}{set random seed $r$ \hl{@xinyi: here r is used by the later random variable, should we use anothoer notation or just throw away this seed thing?}}
\For{$i\leftarrow 1$ to $d$}{\label{forins}
$r_i\sim \small{\textit{Gamma}(2,1)}$, $c_i\sim \small{\textit{Gamma}(2,1)}$, $\beta_i \sim \small{\textit{Uniform}(0,1)}$\;
$t_i\leftarrow \lfloor p\frac{\log u_i}{r_i} + \beta_i \rfloor$\;
$a_i \leftarrow \log(c_i) - r_i (t_i + 1 - \beta_i)$\;
}
\Output{$i^{*} \leftarrow \textit{arg}\min_i a_i$, $t^{*} \leftarrow t_{i^{*}}$}
\caption{Consistent weighted sampling (CWS) for one data vector and one hash sample.}
\label{alg:cws}
\end{algorithm}

The output is a tuple $(i_x^*,t_x^*)$ for data vector $x$. Using the same random numbers ($r,c$ and $\beta$) for another vector $y$, it holds that
\begin{align}
    Pr(i_x^*=i_y^*,t_x^*=t_y^*)=pGMM(x,y;p).
\end{align}
To obtain $m$ hash samples, we repeat the process for $m$ times with $m$ sets of independent random numbers (i.e., $m$ random seeds).

\vspace{0.1in}
\noindent\textbf{Contributions.} The main contributions of this work include:
\begin{itemize}
    \item We propose the framework of {\em consecutive consistent weighted sampling} (CCWS)  as the cohort-building algorithm which achieves $K$-anonymity as well as good scalability ($O(n)$ complexity). Our  approach combines the spirit of hierarchical clustering and CWS hashing. Instead of doing a one-pass assignment, we split the cohorts consecutively/hierarchically using the hash values for a good control on the cohort sizes.

\vspace{0.1in}

    \item We evaluate CCWS on a dataset with > 70M LinkedIn users and LinkedIn ads campaigns. Compared with the well-known SimHash method~\cite{FLoC2020cohort}, CCWS could improve the macro-recall from 0.699 to 0.844, and the micro-recall from 0.077 to 0.254.
\end{itemize}

\section{Cohort Building with CCWS}

We now formally define the problem. Suppose we have $n$ users $U = \{u_1, ...,u_n\}$, where each user is represented as a vector of length $d$: $u_i \in \mathbb{R}^d$. In industrial use cases, $n$ can be hundreds of millions or billions, and $d$ can also be millions of sparse features. Our goal is to assign each member one cohort ID such that each group reserves $K$-anonymity. Ideally, similar members should be assigned to the same cohort ID. Each cohort $c$ will also be associated with a cohort identity, represented by a vector of length $d$: $c\in \mathbb{R}^d$.

% \section{Background}
% Minwise hashing is the Locality Sensitive Hash for Jaccard Similarity. We introduce the case for binary data vectors in minwise hashing, where each data vector can be viewed as a set. Given a set  $x\in\Omega$, the minwise hashing applies a random permutation: $\pi: \Omega \rightarrow \Omega$ on $x$ and stores only the minimum value after the permutation mapping. Formally,
% 	$$h(x) = \min(\pi(x))$$

% Given sets $x, y$, it can be shown that the probability of collisions is the jaccard similarity between the two sets:
% $$Pr(h(x) = h(y)) = \frac{|x\cap y|}{|x \cup y|}$$

% Consistent Weighted Sampling (CWS) generalizes minwise hashing (which applies to binary data) to positive real-valued vectors. Given two data vectors  $u, v\in\mathbb{R}^D$, and apply \autoref{alg:cws}, the probability of collision is equivalent to its pGMM (p-generalized min max) kernel,
% $$\textit{Pr} \big(\textit{CWS}\left(u, p\right) = CWS\left(v, p\right)\big) = pGMM(u,v; p)$$
% where
% $pGMM(u,v; p) = \frac{\sum_{i=1}^D \min(u_i, v_i)^p}{\sum_{i=1}^D \max(u_i, v_i)^p}$. If $p = 1$ and $u, v$ are binary vectors, then pGMM is equivalent to jaccard similarity, and CWS is equivalent to minwise hashing.

\begin{algorithm}

\LinesNumbered
% \SetKwData{Left}{left}\SetKwData{This}{this}\SetKwData{Up}{up}
% \SetKwFunction{Union}{Union}\SetKwFunction{FindCompress}{FindCompress}
%\SetKwInOut{Input}{Input}
\SetKwInput{Input}{Input}
\SetKwInOut{Output}{Output}
\SetKwData{cohorts}{cohorts}
\SetKwData{cid}{cid}
\SetKwData{cUsers}{cUsers}
\SetKwData{u}{u}
\SetKwData{cwsCnt}{cwsCnt}
\SetKwData{cwsUsers}{cwsUsers}

\SetKwData{maxCnt}{maxCnt}
\SetKwData{maxCWS}{maxCWS}
\SetKwData{hash}{hash}
\SetKwData{splitCohort}{splitUsers}
\SetKwData{r}{r}

\Input{Users $U = \{u_1, u_2,\cdots\}$, pGMM kernel tunable parameter $p$, max iteration $T$, privacy budget $K$ to preserve $K$-anonymity}
\textit{// cohorts is a mapping from cohort ID to members mapping; cohort ID is generated by hashing all the users in the cohort}\\
\cohorts $\leftarrow \{sha256(U): U\}$\\
\For{$t\leftarrow 1$ to $T$}{
    \For{\cid, \cUsers in \cohorts}{
    \If{$|$\cUsers$| < 2 * K$}{
        continue\;
    }
     \cwsUsers $\leftarrow \{\}$  \quad \textit{ //  hash to users map} \;
     \cwsCnt $\leftarrow \{\}$ \quad \textit{ //  hash to count map} \;
        \For{\u in \cUsers}{
            $m$ $\leftarrow$ generate random seed \;
            \hash = CWS(\u, $p$, $m$) \quad  \textit{// use Algorithm 1}\;
            \cwsCnt$[$\hash$]$ += 1 \;
            Add \u to \cwsUsers$[$\hash$]$\;
        }
    \maxCWS, \maxCnt $\leftarrow$ $\max(\cwsCnt)$ \ \ \textit{//max count}\;
    \If{\maxCnt >= $K$ and $|$\cUsers$|$ - \maxCnt >= $K$ }{
    \textit{// Split only when $K$-anonymity can be enforced}
        \splitCohort $\leftarrow $ \hash[\maxCWS]\;
        add $\{sha256($\splitCohort) $\rightarrow$ \splitCohort $\}$ \;
        add $\{sha256(U \backslash \,$\splitCohort) $\rightarrow  U \backslash$ \splitCohort $\}$ \;
        delete \cid in \cohorts\;
        }
    %\EndIf
    }
}
\Output{\cohorts}
\caption{The proposed consecutive consistent weighted sampling (CCWS) for cohort building. }
\label{alg:main}
\end{algorithm}

\subsection{The Proposed CCWS Algorithm}
The  algorithm is presented in Algorithm~\ref{alg:main}. At initialization, we start by assigning all users into one initial cohort. In each iteration, for each intermediate cohort (\texttt{cid}, \texttt{cUsers}), we do the following:
\begin{enumerate}
    \item[1.] (Line 10 - Line 15) We apply CWS to generate one hash value for every user within the cohort (with same seed). We also create a map (\texttt{cwsCnt}) that stores the count of distinct hash values within the cohort.

    \item[2.] (Line 17 - Line 21) We find the maximal  count \texttt{maxCnt} and its corresponding hash value \texttt{maxCWS}. If $\texttt{maxCnt}>K$ and the number of remaining users with hash value different from \texttt{maxCWS} is also  $>K$, we split the cohort into two according to the hash \texttt{maxCWS}. This ensures  the two new cohorts both have at least $K$ members. Otherwise, at the current iteration we do not split that cohort, which might still be split in later iterations as long as the condition on Line 17 is satisfied.

    \item[3.] (Line 5 - Line 7) If an intermediate cohort has less than $2 \times K$ members, there is no way to split it into two cohorts both containing more than $K$ users. In this case, this cohort will not be split anymore and is included in the final cohort set.
\end{enumerate}

The procedure repeats until no cohorts can be further split. The final set of cohorts (i) satisfies $K$-anonymity as all the user groups have more than $K$ members;  (ii) is ``similarity-preserving'' as users within each group tend to have high pGMM similarity (\ref{def:pGMM}). Note that, we only use the first integer hash value $i^*$ in CCWS (Algorithm~\ref{alg:cws}), because prior research on CWS reported that $Pr(i_x^*=i_y^*)\approx pGMM(x,y;p)$ and the approximation is very good~\cite{li2017tunable,li2017theory, li2021consistent, li2022gcwsnet}, which is called ``0-bit CWS''. In other words, only using the first hash value will not undermine the utility at least empirically.

The algorithm is flexible in real-world use cases. For example, prior knowledge can be leveraged easily. If there are important features (e.g., geo-location) that each cohort must have, then we can set the initial cohorts as users with/without the important features, as opposed to starting with one initial cohort with all users. Moreover, the (quantified) importance of different features can also be integrated with CCWS smoothly. Before Line 12, we can optionally multiply a feature importance vector by $u$ to emphasize more important features. The feature weights can be related to business metrics, legal regulations, etc., making CCWS a convenient cohort building tool to meet different needs in industrial practice.

\subsection{Complexity of CCWS}

By using the hash tables, in each iteration, CCWS only needs to generate one hash value for all the data points, which takes $O(n)$ times. Thus, the total time complexity of CCWS is $O(Tn)$, where $T$ is the number of iterations, and $n$ is the number of users. Practically, we find that $T=1000$ is enough for $n=100M$ users. Since CCWS effectively only builds one hash table, the memory complexity is simply $O(n)$. Therefore, the computational complexity and the memory complexity of CCWS are both linear in $n$. As a result, CCWS is substantially more efficient than the classic clustering algorithms mentioned in Section~\ref{sec:intro} in terms of both speed and space.

There have been a series of works on improving the efficiency of minwise hashing and consistent weighted sampling. These include ``circular minwise hashing'' (C-MinHash)~\cite{li2022c}, ``binwise CWS'' (BCWS)~\cite{li2019re},  ``densified one permutation hashing`` (OPH)~\cite{shrivastava2014improved}, etc.

\subsection{Hash-and-Sort as Strong Baselines}  \label{sec:sorting method}

One might ask whether we can directly set the cohorts as the buckets in the hash table~\cite{indyk1998approximate,shrivastava2012fast}, which might lead to a more straightforward algorithm. This naive bucketing strategy, however, exhibits the same issue as DBSCAN and $k$-means, in that we cannot directly control the cohort (bucket) size to maintain $K$-anonymity.

To fix this issue, one approach is to use a ``hash-and-sort'' type strategy~\cite{ravichandran2005randomized}, which has been adopted in industry. In~\cite{FLoC2020cohort}, they first generate a  bit vector using SignRP (SimHash) for each user.  Denote the bit vectors as $v_1, .., v_n$ for all the users. Then, we sort them by lexicographical order and obtain a sorted list $v_{(1)}\leq \cdots \leq v_{(n)}$. Finally, we construct cohorts by grouping every $K$ items from low to high consecutively, i.e., $H_1=\{v_{(1)},...,v_{(K)}\}$, $H_2=\{v_{(K+1)},...,v_{(2K)}\}$, etc. This approach is similar to the bucketing strategy, but enforces a fixed group size. Of course, CWS or MinHash can also be applied in this strategy to replace SignRP. That is, we  lexicographically sort the hash vectors consisting of  values from CWS or MinHash, instead of those from SignRP. Our experiments will   report the results of the hash-and-sort strategy applied to those hashing methods.

\section{Experiments}

We evaluate cohort-building algorithms on a dataset with more than 70M LinkedIn users. Each user has a weighted non-negative feature vector of length $d$ about 200,000. For the minimal cohort size, we use $K=20$ in all the experiments, as \cite{FLoC2020} has shown that the re-identification risk, as $K$ increases, becomes almost flat (and very low) as long as $K$ reaches 20. At the time of writing this paper, we also noticed that Google claimed to adopt $K=50$ in their latest updated privacy sandbox FLEDGE~\cite{Fledge2023K} on February 9th, 2023.

\vspace{0.05in}
\noindent We compare the following cohort-building algorithms:

\begin{itemize}
    \item \textbf{Random Grouping:} we uniformly randomly assign the users into 3 million cohorts, as a naive baseline.
    \item \textbf{SignRP/MinHash/CWS:} we implement the ``hash-and-sort'' strategy described in Section~\ref{sec:sorting method}, using SignRP, MinHash and CWS as the underlying hashing method, respectively. For all these methods, we search over the length of the hash vectors in $\{50, 75, 100\}$. For MinHash, we simply binarize the data by treating all non-zero entries to $1$.

    \item \textbf{CCWS:} We run our proposed Algorithm~\ref{alg:main} with max $T=1000$ iterations. We tune the parameter $p$ in the pGMM kernel~\cite{li2017tunable} on a grid from 0.5 to 1.5 spaced at 0.1.

\end{itemize}

\vspace{-0.05in}

\subsection{Evaluation Metrics}
We use LinkedIn ads campaigns to evaluate the cohorts. We define the campaign level metrics as follows. For each campaign $c$:
\begin{itemize}
    \item True Positives ($TP_{c}$): the number of users who are matched by campaign $c$, and his/her cohort is also matched by $c$;
    \item False Positives ($FP_{c}$): the number of users who are not matched by campaign $c$, and his/her cohort is matched by $c$;
    \item False Negatives ($FN_{c}$): the number of users who are matched by campaign $c$, and his/her cohort is not matched by $c$.
\end{itemize}
Given a set of campaigns $C$, the macro-recall and micro-recall are
\begin{align*}
\textit{macro-recall} &= \frac{\sum_{c\in C}| TP_c|}{\sum_{c \in C}|TP| + \sum_{c \in C}|FN_c|},\\
\textit{micro-recall} &= \frac{1}{|C|}\sum_{c\in C}\frac{\left|TP_{c}\right|}{\left|TP_{c}\right| + \left|FN_{c}\right|}.
\end{align*}

\vspace{-0.2in}
\subsection{Results}
\vspace{-0.05in}

\subsubsection{Macro-recall and micro-recall}

\begin{table}[h]

\vspace{-0.03in}

    \centering
    \caption{Evaluations of cohort-building algorithms.\vspace{-0.15in}}
    \begin{tabular}{c|cc}\hline
        Method & micro-recall & macro-recall \\\hline
        Random Grouping & 0.004 & 0.044\\
        MinHash & 0.064 &  0.621\\
        SignRP & 0.077 & 0.699\\
        CWS & 0.082 & 0.721 \\
        CCWS & \textbf{0.254} & \textbf{0.844} \\
         \hline
    \end{tabular}
    \label{tab:recall}\vspace{-0.12in}
\end{table}

From \autoref{tab:recall}, among the three ``hash-and-short'' methods, CWS outperforms SignRP and MinHash. This is not surprising, as \cite{shrivastava2014defense} already demonstrated the disadvantage of SignRP on sparse data. Here MinHash only used binarized feature hence it did not perform as well as SignRP for our task.  This suggests that feature weights are very helpful for cohort-building.

The proposed CCWS significantly outperforms all other methods. In particular, CCWS exhibits a huge improvement over CWS on micro-recall. The plausible reason is that the ```hash-and-sort''' strategy is sub-optimal when grouping users, so that the users within cohorts are less similar than those in CCWS.
Also, note that, even though CCWS can already achieve a $>0.8$ macro-recall, the method has a relatively low micro-recall.  We believe this is because macro-recall is dominated by campaigns whose audience sizes ($TP + FP$) are large. On campaigns with smaller audience sizes (a.k.a., campaigns with more detailed targeting criterion), there is still ample room for improvement as the future research study.

\subsubsection{Cohort size distribution}

\begin{figure}[ht]

\vspace{-0.1in}

{\hspace*{-.2in}
\begin{minipage}{.245\textwidth}
  %\hspace*{-.1in}
  \includegraphics[width=1.03\linewidth]{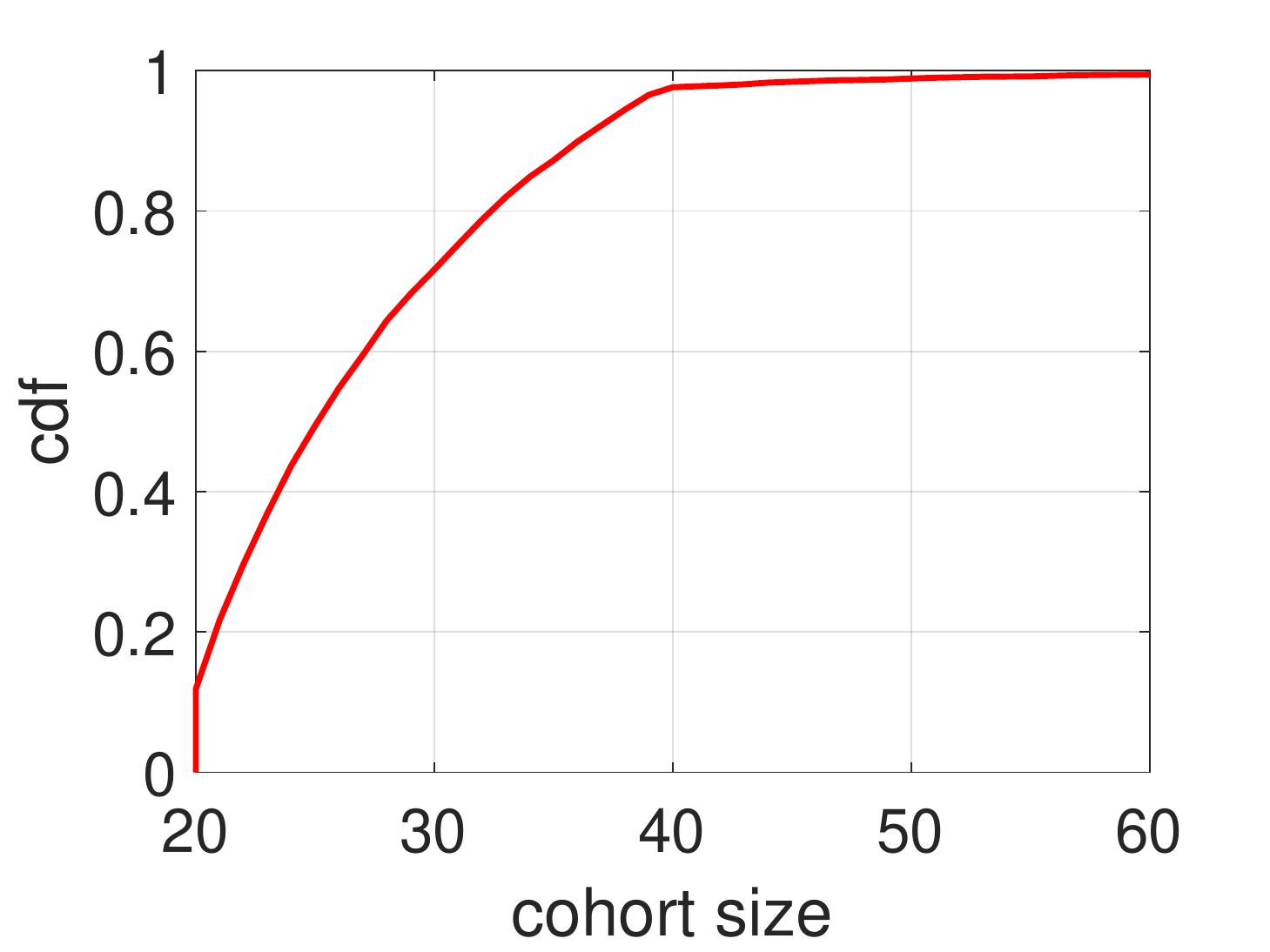}

  \vspace{-0.15in}
  %\hspace*{1in}

  \captionof{figure}{Empirical cumulative distribution function (cdf) of  cohort sizes using CCWS.}
  \label{fig:cdf}
\end{minipage}%
\hspace*{.1in}
\begin{minipage}{.245\textwidth}
%\vspace*{-0.0in}
  \includegraphics[width=1.03\linewidth]{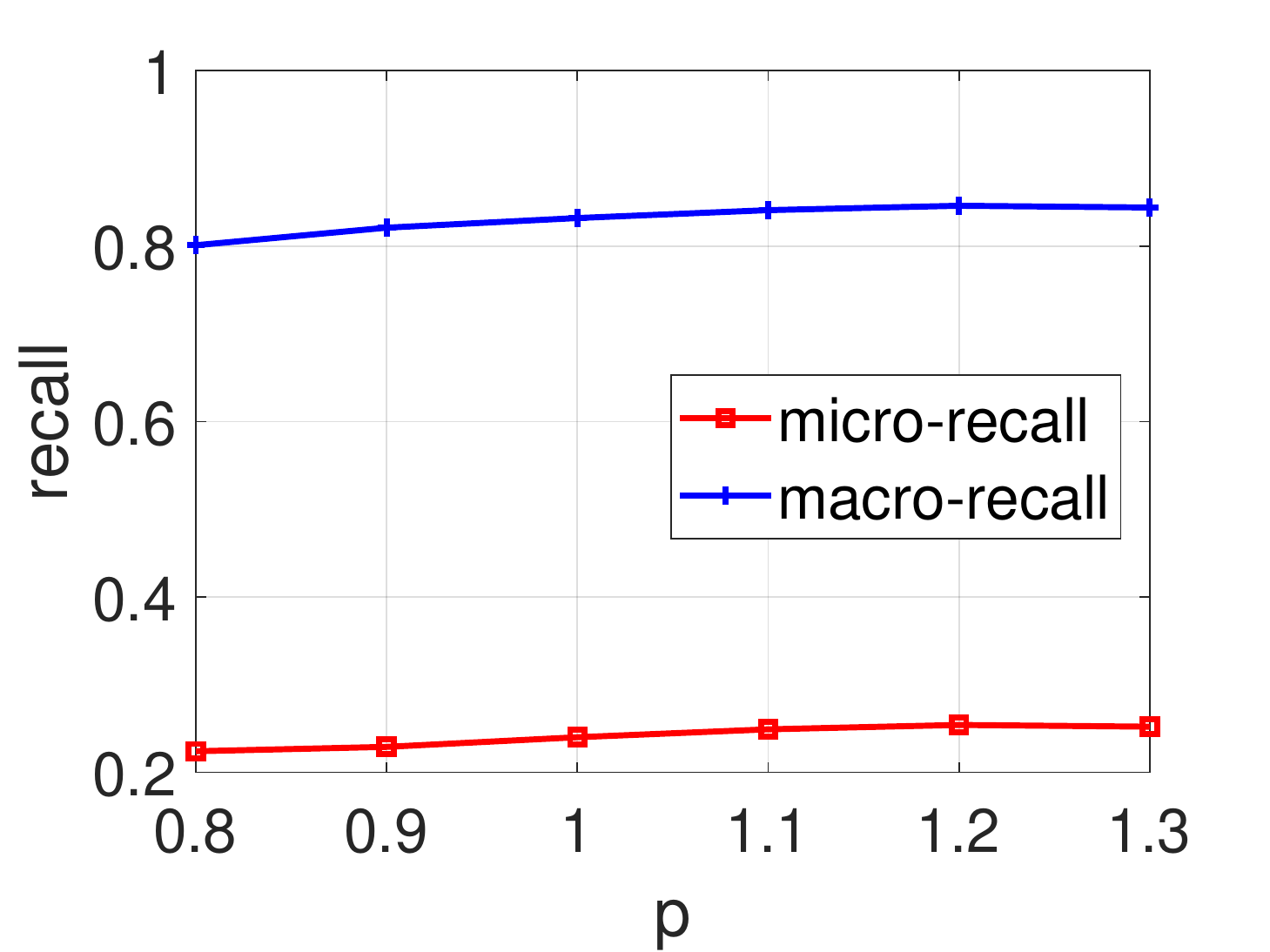}

 \vspace{-0.15in}

  \captionof{figure}{CCWS: tunable parameter $p$ vs. micro-recall and macro-recall metrics.}
  \label{fig:p}
\end{minipage}
}
\vspace{-0.15in}
\end{figure}

In \autoref{fig:cdf}, we visualize the (empirical) cumulative distribution function (cdf) of cohort size distribution generated by CCWS. We see that the vast majority (>95\%) of cohorts have size between 20-40. The 99-th percentile size for CCWS is 57. This suggests that $T=1000$ rounds of iterations might be sufficient for generating good cohorts for this 70M scale dataset. This also illustrates that, CCWS not only  achieves better utility than other methods, but also maintains stronger privacy in the sense that CCWS allows the cohort size to be greater than $K=20$.

% \begin{figure}[htb]

% \centering

% \includegraphics[width=2.5in]{plots/recall_vs_p.eps}

% \vspace{-0.1in}

%     \caption{CCWS: parameter $p$ vs. recall metrics.}
%     \label{fig:p}
% \end{figure}

\subsubsection{Effects of $pGMM$ tunable parameter $p$}
Recall from Eq.~\eqref{def:pGMM} and Algorithm~\ref{alg:cws}  that CWS has a parameter $p$, which essentially tunes the weighted Jaccard Similarity. The impact of $p$ is presented in~\autoref{fig:p}, which shows that $p=1.2$ is most effective in both micro-recall and macro-recall. In general, we see that the impact of $p$ is not very significant in the neighborhood of 1. In other words,  CCWS is fairly robust against hyper-parameter tuning.

\section{Conclusion}

With the privacy landscape shift in the digital advertising world, building cohort-based identity is  increasingly important for  user-privacy and personalized ads.
In this regard, we present CCWS (consecutive CWS)  for cohort building to achieve $K$-anonymity and demonstrate that CCWS significantly outperforms three popular hashing methods including MinHash, SignRP, and (vanilla) CWS on LinkedIn user members and campaigns. We implement MinHash, SignRP, and CWS in a ``hash-and-sort'' fashion so that we can conveniently control the cohort size. CCWS integrates the advantages of both CWS and hierarchical clustering and it hence achieves the best performance. Among other methods, CWS considerably outperforms SignRP and MinHash in our application, which uses high-dimensional sparse (and non-zero) features. The popular SignRP (a.k.a. SimHash) has the disadvantage in sparse data as shown in~\cite{shrivastava2014defense}. On the other hand, the non-zero feature values carry useful information and thus MinHash (which uses binarized data) does not perform  as well as SignRP. We hope our study can be interesting to both the industry practice and academia research.

\newpage
\section*{Company Portrait}
\noindent\textbf{About LinkedIn:}  Founded in 2003, LinkedIn connects the world's professionals to make them more productive and successful. With more than 850 million members worldwide, including executives from every Fortune 500 company, LinkedIn is the world's largest professional network. The company has a diversified business model with revenue coming from Talent Solutions, Marketing Solutions, Sales Solutions and Premium Subscriptions products. Headquartered in Silicon Valley, LinkedIn has offices across the globe. \url{https://www.linkedin.com/company/linkedin/about/}

\section*{Presenter Bio}
\noindent\textbf{Xinyi Zheng} is a Software Engineer at LinkedIn. She received her M.S. in Computer Science from the Carnegie Mellon University, and her B.S. in Computer Science and Mathematics from the University of Michigan. Her research interests lie in recommender systems and graph mining. \textbf{Ping Li} (\url{https://pltrees.github.io}) is a Distinguished Engineer at LinkedIn. He obtained his PhD in Statistics, MS in Computer Science, and MS in Electrical Engineering, from the Stanford University. He also received two master's degrees from the University of Washington (Seattle). Before joining LinkedIn, Ping Li was the Deputy Dean of Baidu Research after a memorable academic career at Rutgers University and Cornell University. During the time as an assistant professor, Ping Li received  the Young Investigator Award from the Office of Naval Research (ONR-YIP) and the Young Investigator Award from the Air Force Office of Scientific Research (AFOSR-YIP). Ping Li's research interests include boosted trees (\url{https://github.com/pltrees/abcboost}), approximate near neighbor (ANN) search,  fast neural ranking, embedding-based retrieval (EBR), generative models, big models for search/ranking/feeds/advertising, AI model security, federated learning, and differenitial privacy.

\balance
\bibliographystyle{ACM-Reference-Format}
\bibliography{refs_scholar}

\end{document}